\documentclass[epj]{svjour}
\usepackage{epsf,amsmath,amssymb}
\usepackage{graphicx}

\newcommand{\bbr}{{\boldsymbol r}}

\begin{document}

\title{ Melting and evaporation transitions in small Al clusters: 
  canonical Monte-Carlo simulations}
\author{ Ralph Werner }   
\institute{Institut f\"ur Theorie der Kondensierten Materie, Universit\"at
Karlsruhe, D-76128 Karlsruhe, Germany }

\date{Version: \today}

\abstract{ A dimer of bound atoms cannot melt, only dissociate. Bulk
  metals show a well defined first order transition between their
  solid and liquid phases. The appearance of the melting transition is
  explored for increasing clusters sizes via the signatures in the
  specific heat and the root mean square of the bond lengths 
  $\delta_{\rm B}$ (Berry parameter) by means of Monte-Carlo
  simulations of Al clusters modelled by Gupta potentials. Clear
  signatures of a melting transition appear for $N\sim 6$
  atoms. Closed-shell effects are shown for clusters with up to 56
  atoms. The melting transition is compared in detail with the
  dissociation transition, which induces a second and possibly much
  larger local maximum in the specific heat at higher
  temperatures. Larger clusters are shown to fragment into dimers and
  trimers, which in turn dissociate at higher temperatures.
}

\maketitle

\section{Introduction}

The properties of small metal clusters have
enjoyed a large interest over the past years. Their technical
application in catalysis stems from the large surface-to-volume ratio
while their properties differ from those of the bulk material raising
the fundamental question about the statistical mechanics of finite
systems.  

The melting process of small clusters has early on been identified as
the onset of isomer fluctuations \cite{AB86,JBB86}. More recent
investigations on Ni$_{13-x}$Al$_x$ alloy clusters in the
microcanonical ensemble \cite{KJ97} show the relation between isomer
fluctuations and the increase in entropy across the melting
transition. The onset of the melting transition is marked by the
fluctuations into the lowest energy isomer configurations which are
measured by the root mean square bond length fluctuations
\cite{BBDJ88} sometimes referred to as the Berry parameter
\cite{ZKBB02}. The phase space occupied by those fluctuations is small
for small clusters resulting in a maximum in the specific heat at
somewhat higher temperatures. The maximum in the specific heat in turn
is determined by the onset of isomer fluctuations occupying a
sufficiently large phase space fraction. A detailed overview of the
increase of phase space with increasing particle number and the
classification of isomers in terms of potential energy surfaces is
given in Ref.\ \cite{WDM+00}. Molecular Dynamics (MD) investigations
\cite{EAT91} of the melting transition of larger Au$_{N}$ clusters
with $100 < N < 1000$ show that the bulk limit is gradually attained
in agreement with experimental \cite{ESZ+00} findings.

Recently the empirical investigation of the melting of small Sn
\cite{SJ00} and Ga \cite{BBS+03} clusters has revealed a possible
stability of the solid phase of the particles beyond the melting
temperature of the bulk material. This result was interpreted as a
consequence of the specific rigid ground state structure of the
clusters and found support in microcanonical MD calculations for C,
Si, Ge, and Sn clusters \cite{LWH00} as well as for isokinetic MD
investigations of Sn$_{10}$ particles \cite{JKB02}.

In metals the contribution from the conduction electrons to the
binding energy has to be modelled by many-body potentials
\cite{Gupt81,TMB83}, which are numerically more involved than the
thoroughly investigated Lennard-Jones systems
\cite{WDM+00,HA87,Fran01}. A prominent example is the Gupta potential 
(GP) \cite{Gupt81}, which can be derived in the second moment
approximation from a tight binding model \cite{ZLT91} and which
correctly describes the surface contraction observed in metals: 
\begin{equation}\label{Gupta}
V(\{r_{ij}\}) = \sum_{i}^N\left[\sum_{j\neq i}^N
A e^{-p\, \overline{r}_{ij}} - 
\sqrt{\sum_{j\neq i} \xi^2 e^{-2q\, \overline{r}_{ij}}}\right]\,.
\end{equation}
Here $N$ is the number of atoms, $i$ and $j$ are atom labels,
$\overline{r}_{ij} = r_{ij}/r_0 - 1$, and $r_{ij} = |\bbr_i - \bbr_j|$
is the  modulus of the distance between two atoms at positions
$\bbr_i$ and $\bbr_j$. The parameters have been determined by fitting
the experimental bulk lattice parameters and elastic moduli
\cite{CR93} as $A = 0.1221$ eV, $\xi = 1.316$ eV, $p = 8.612$, and $q
= 2.516$ for Al. Distances are measured in units of the bulk first
neighbour distance $r_0 = 2.864$ \AA.

The present paper aims to shed light on how the melting transition
evolves in the limit of small clusters. The method is a MC simulation
in the canonical ensemble. A standard Metropolis algorithm is employed
\cite{AT89,WDS+01} with an update after each random displacement of an
atom within an interval $[0,d_{\rm max}]$ in all spatial
dimensions. $d_{\rm max}$ is set to yield an MC acceptance rate of 50
to 60 \%. The resulting temperature dependence is roughly $d_{\rm max}
\propto \sqrt{T}$. The boundary conditions are imposed by a hard wall
cube with linear dimension $L$. Runs are performed with sampling rates
(SR) of up to $8\times10^7$ steps per temperature and atom. The 
fluctuations on the curves shown in the paper are a measure of the
statistical error and appear near phase transition because of the
usual critical slowing down. The ground state energies and
configurations obtained within this method are in good agreement with
earlier results \cite{KJ97,TJW00}. The ground state configurations of
the clusters discussed herein have the same symmetries as those of the
9-6 Sutton-Chen potentials \cite{DW98}.

An observable commonly studied in the context of melting transitions is
the Berry parameter \cite{BBDJ88}
\begin{equation}\label{Berry}
\delta_{\rm B} = \frac{1}{N(N-1)}\sum_{i,j\neq i}
    \sqrt{\langle r_{ij}^2 \rangle - \langle r_{ij} \rangle^2}\
         \langle r_{ij} \rangle^{-1}\,,
\end{equation}
where the brackets denote the thermodynamic average in the canonical
ensemble. The parameter Eq.\ (\ref{Berry}) measures the root mean
square of the distance between two atoms averaged over all pairs. Even
short isomer fluctuations with a subsequent return to the ground state
can leave the cluster reordered, i.e., a previously nearest neighbour
pair $r_{ij}$ may become a second-  or third-neighbour pair after the
fluctuation. Such a reordering leads to a notable increase in
$\delta_{\rm B}$ allowing for the clear identification of a melting
transition. Note that the Lindemann criterion of melting, which
measures the atomic fluctuations with respect to their equilibrium
positions, is usually employed for bulk systems but is less well
suited for cluster systems \cite{ZKBB02}.

The second observable of interest is the specific heat, which can be
obtained from the thermodynamic averages of the potential energy $V$
and its square: 
\begin{equation}\label{C}
\frac{C}{k_{\rm B}} = \frac{1}{Nk_{\rm B}^2T^2}
    \left(\langle V^2 \rangle - \langle V \rangle^2\right) 
    + \frac{3}{2}\,.
\end{equation}
Since the atoms are treated as classical particles the
contribution from the kinetic energy is $C_{\rm kin} = 3/2 k_{\rm B}$
per atom.

\begin{figure}
   \epsfxsize=0.49\textwidth
   \centerline{\epsffile{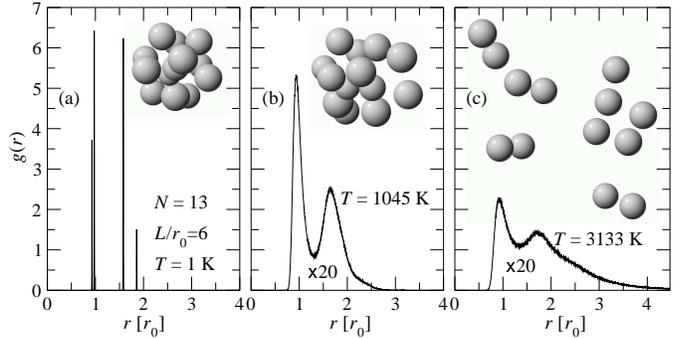}}
   \caption{\label{gui}\sl
     Normalised pair distribution functions and snapshots of an
     Al$_{13}$ cluster in a volume of $(6 r_0)^3$ at temperatures (a)
     $k_{\rm B}T = 10^{-4}$ eV, (b) $k_{\rm B}T = 0.09$ eV, and (c)
     $k_{\rm B}T = 0.27$ eV corresponding to a solid, liquid, and
     dissociated cluster, respectively. The pair distribution
     functions in panel (b) and (c) are enhanced by a factor of 20 for
     better visibility.
   }
   \end{figure}

In order to obtain a more intuitive understanding of the melting
process a real-time visualisation of the simulation has been 
implemented. Figure \ref{gui} illustrates snapshots of an Al$_{13}$
cluster in a volume of $(6 r_0)^3$ at temperatures (a) $T = 0$, (b)
$k_{\rm B}T = 0.1$ eV, and (c) $k_{\rm B}T = 0.4$ eV corresponding to
a solid, liquid, and dissociated cluster, respectively. The graphs
show the corresponding normalised pair distribution functions $g(r)$
(arbitrary units). The ground state is icosahedral which is different
but close to the slightly distorted icosahedral configuration obtained
by first principle calculations \cite{AE99}. Contacts of the clusters
with the walls are rare events and the pressure is negligible in the
solid and liquid phases for sufficiently large volumes, e.g., $L >
4r_0$ for $N=13$.    

\section{Appearance of the melting transition}

\begin{figure}
     \resizebox{0.485\textwidth}{!}{%
       \includegraphics{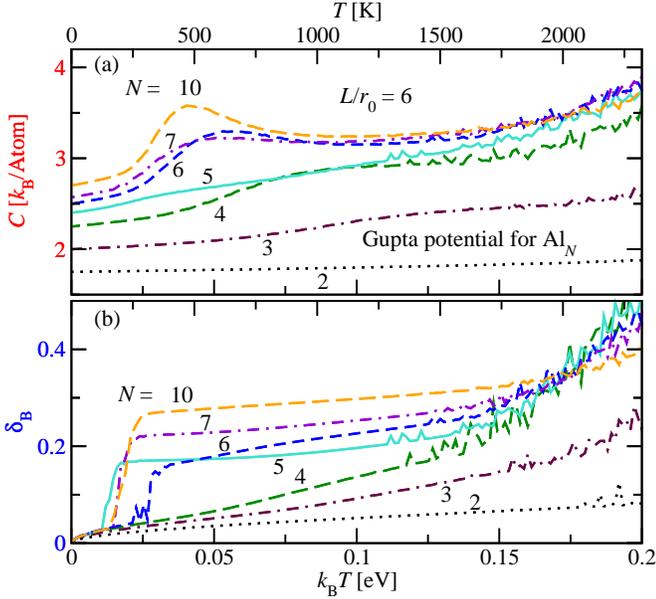}
     }
     \caption{\sl
       Temperature dependence of the specific heat (a) and the Berry
       parameter (b) of Al$_N$ clusters with $N=2,3,4,5,6,7,10$. A
       clear signature of a melting transition is  
       observed for $N\ge6$ in the specific heat and for $N\ge5$ in the
       $\delta_{\rm B}$. SR: $4\times10^7$ steps per 
       temperature and atom. 
     }
     \label{CRofT}
   \end{figure}

A dimer of bound atoms cannot melt, only dissociate. Bulk metals show
a well defined first order transition between their solid and liquid
phases. This phase transition is accompanied by a divergence in the
temperature dependence of the specific heat indicating the increase in
entropy and the associated latent heat as well as by a discontinuous
jump in the Berry parameter Eq.\ (\ref{Berry}). Figure \ref{CRofT}
shows how both signatures evolve [specific heat (a) and Berry
parameter (b)] as the particle number is increased from 2 to 10 atoms.

Al$_5$ is the smallest cluster with inequivalent bonds in its ground
state configuration resulting in the abrupt increase in $\delta_{\rm
  B}$ once isomer fluctuations set in at around $k_{\rm B}T \sim
0.013$ eV or $T \sim 150$ K. For Al$_N$ with $N\ge 6$ the increase in
entropy is sufficiently large around a specific temperature to lead to
a local maximum in the specific heat. As observed previously
\cite{KJ97}, the isomer fluctuations leading to the jump in
$\delta_{\rm B}$ occur at lower temperatures than the maximum in the
specific heat. The temperature of the discontinuity in $\delta_{\rm
  B}$ depends on the energy barrier between the ground state
configuration and the lowest energy isomers \cite{WDM+00}.

Note that the presence of well defined signatures of the melting
transition in the specific heat for clusters with $N < 10$ is somewhat
unexpected since the experimentally investigated Na$_N$ clusters
\cite{SDHH03} appear to show no feature in the caloric curves for $N
< 100$ and for Lennard-Jones clusters \cite{Fran01} the corresponding
signature disappears for $N < 30$.

It is remarkable that the absence of signatures of a well defined
melting transition does not imply that the clusters remain solid up to
higher temperatures. For example, the real time visualisation of an
Al$_4$ particle (within the MC/GP approach, see Fig.\ \ref{gui} for an
illustration) at $T = 0.1$ eV/$k_{\rm B}$ $\sim 1160$ K reveals that
the cluster fluctuates out of its tetrahedral ground state into almost
planar configurations and certainly cannot be considered solid. The
small size of the phase space is responsible for the absence of a
signature of the melting in the specific heat while the equivalence of 
all bonds in the ground state configuration result in a featureless
$\delta_{\rm B}$.

In this sense all clusters investigated here with $N\ge 4$ are in a
liquid state already at temperatures below the bulk melting
temperatures of Al with $T_{\rm bulk} = 933$ K $\sim 0.08$ eV/$k_{\rm
  B}$. For $N=2$ it is not possible to distinguish between a liquid
and a solid phase. For $N=3$ the triangular structure is stable
against fluctuations into collinear isomers up to $k_{\rm B}T \approx
0.15$ eV.

The critical value of the Berry parameter at the melting transition
\cite{ZKBB02} was determined as $\delta_{\rm B} \sim 0.03 -
0.05$, which is referred to as the modified Lindemann criterion. These
numbers are consistent with the present observation for clusters with
$N \ge 5$, where at the melting transition a jump occurs from values
of $\delta_{\rm B} \sim 0.03 - 0.05$ to values of $\delta_{\rm B} >
1.5$ \cite{AB86,JBB86,KJ97,WDS+01,BB01}. For smaller clusters with $N
= 3$ and $4$ the temperature dependence of the Berry parameter is
featureless. For $N=4$ at $T = 0.1$ eV/$k_{\rm B}$ and for $N=3$
at $T=0.15$ eV/$k_{\rm B}$ the real time rendering of these
clusters shows that they cannot be considered solid any more. At those
temperatures the Berry parameter is $\delta_{\rm  B} \approx
0.1$. These findings suggest that a more general sufficient criterion
for clusters of all sizes not to be considered solid any more is
$\delta_{\rm B} \ge 0.1$. The latter is close to the value given for
the Lindemann criterion \cite{ZKBB02}.  

Multi-step melting \cite{BB01,KB94} and isomer fluctuations
\cite{JKB02} involving reordered atomic arrangements in the cluster
are also consistent with that criterion since in both cases at least a
group of atoms does not remain located at their ground state positions
when $\delta_{\rm B} \ge 0.1$. 

Both specific heat and Berry parameter in Fig.\ (\ref{CRofT})
show an increase for temperatures above $\sim 1800$ K accompanied by
an increase in fluctuations due to statistical errors. As will be
discussed further below in detail these are signatures of the
dissociation transition.

\section{Closed-shell effects}\label{sectionClosed}

\begin{figure}
     \epsfclipon
     \epsfxsize=0.49\textwidth
     \centerline{\epsffile{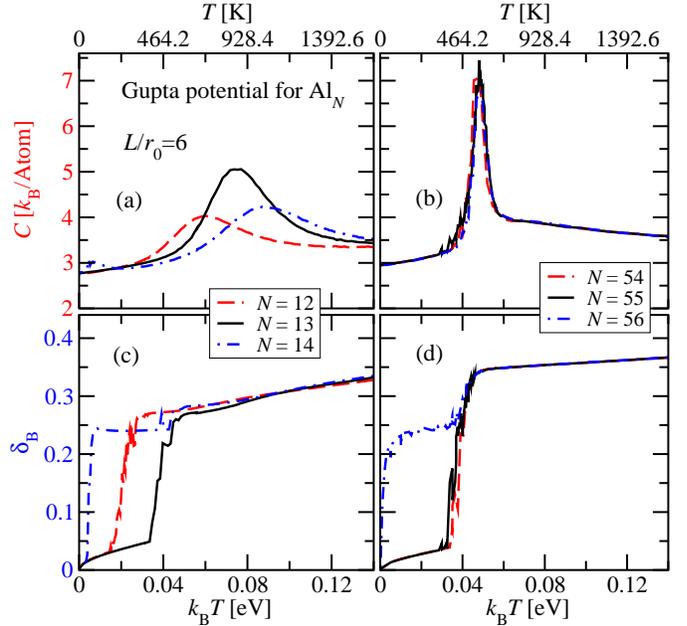}}
     \caption{\label{CRofTshell}\sl
       Temperature dependence of the melting transition for clusters of
       size $N = 12, 13, 14$ in (a)+(c) and $N = 54, 55, 56$ in
       (b)+(d). (a) and (b): specific heat. (c) and (d): Berry
       parameter. Volume $L^3/r_0^3 = 6^3$. SR: $4$ to 
       $8\times10^7$ steps [(a)+(c)] and $1$ to
       $2\times10^7$ steps [(b)+(d)] per temperature and atom.   
     }
   \end{figure}

The influence of closed shells on the cohesive energies of metal
clusters \cite{DW98,AE99,TJW00,WDM+00} and their melting points
\cite{ESZ+00,BB01,WDS+01} has been a focus of research for quite some
time. A closed-shell cluster has a large gap to the first excited
isomer while adding or removing an atom leads to a number of
degenerate ground state configurations separated by a potential
barrier. This manifests itself in a smoother specific heat anomaly as
well as in a jump in the Berry parameter at much lower temperatures as
compared to the closed-shell counterpart. This is shown in Fig.\
\ref{CRofTshell}(a) and (c) for the sets of $N = 12, 13, 14$ and in
Fig.\ \ref{CRofTshell}(b) and (d) for the sets of $N = 54, 55, 56$
atoms. The upper panels (a) and (b) show the specific heat, the lower
panels (c) and (d) the Berry parameter.

Notably the 14 and 56 atom clusters have a very low barrier between
isomers. The real time visualisation reveals that these fluctuations
occur in Al$_{14}$ not only by jumps of the 14$^{\rm th}$ atom on the
surface of the Al$_{13}$ icosahedron but by absorption of the 14$^{\rm
  th}$ atom into the outer shell and simultaneous pushing of another
atom onto the surface. In Al$_{56}$ the 56$^{\rm th}$ atom is absorbed
into the outer shell even in the ground state configuration (see
Sutton-Chen 9-6 in Ref.\ \cite{DW98}). This leads to the large jump in
$\delta_{\rm B}$ at very low temperatures [dash-dotted lines in Fig.\
\ref{CRofTshell}(c) and (d)]. In Al$_{14}$ the phase space is
sufficiently large to induce an anomaly in the specific heat
[dash-dotted line in Fig.\ \ref{CRofTshell}(a)] at the same
temperature. Together with the main maximum at higher temperatures
this may be referred to as a two-step melting mechanism
\cite{BB01,KB94,CB92}.  

The narrower specific heat anomaly at the melting transition and the
smaller discrepancy between the maximum of the specific heat
and the jump in the Berry parameter for the larger systems in Fig.\
\ref{CRofTshell}(b) and  (d) illustrate how the thermodynamic limit is
gradually approached as the cluster size is increased
\cite{EAT91,BB01,SKK+97}. Note that for larger clusters less sampling
steps per temperature and atom are required to obtain smooth
curves. The larger phase space of the larger clusters \cite{WDM+00}
results in the better convergence of the observables.

The canonical ensemble as shown for Al$_{13}$ in Fig.\
\ref{CRofTshell}(a) and (b) yields a somewhat lower specific heat 
\cite{BB01} and an onset of the isomer fluctuations at lower
temperatures as compared with the results for the microcanonical
ensemble in Ref.\ \cite{KJ97}. The discrepancies can be attributed to
the energy fluctuations in the canonical ensemble that allow the
potential barriers between different isomers to be overcome  at lower 
temperatures.

\section{Ideal gas limit}

\begin{figure}
     \resizebox{0.485\textwidth}{!}{%
       \includegraphics*[clip=False]{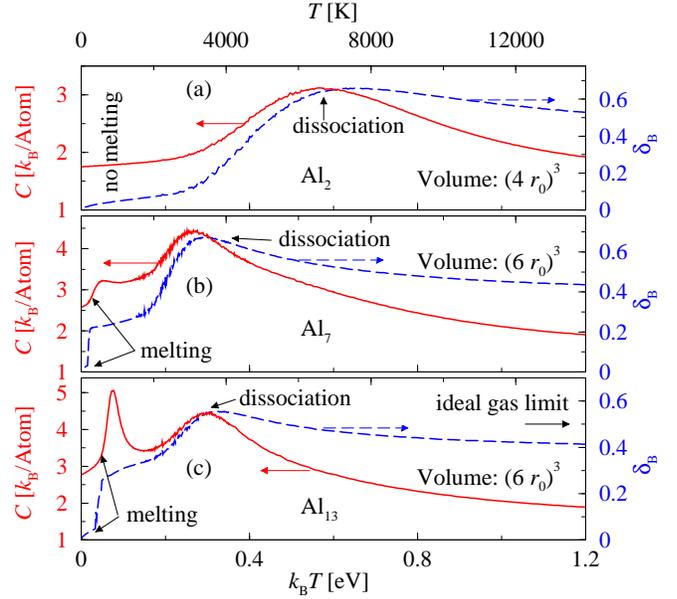}}
     \caption{\label{CRofLargeT}\sl
       Temperature dependence of the specific heat (full line, left scale)
       and the Berry parameter (broken line, right scale) of a 2 (a), 7
       (b), and 13 (c) atom Al cluster. The dissociation anomaly is
       always present while the signature of the melting transition
       evolves with system size. SR: $4\times10^7$ [(a)+(b)] and
       $2\times10^7$ (c) steps per temperature and atom. 
     } 
   \end{figure}

The well defined high temperature limit, where the system has the 
properties of an ideal gas, yields a test of the numerical methods and
is essential for determining the latent heat of the evaporation
transition in Sec.\ \ref{sectionDiss}. Figure \ref{CRofLargeT}
illustrates the large $T$ behaviour of the specific heat (full line,
right scale) and the Berry parameter (broken line, left scale) for
Al$_{2}$ (a), Al$_{7}$ (b), and Al$_{13}$ (c). Since the Boltzmann
weight can be expanded as $\exp\{-V/(k_{\rm B}T)\} = 1 - V/(k_{\rm
  B}T) + {\rm O}(T^{-2})$, the specific heat attains the limit as $C =
\frac{3}{2} k_{\rm B} + a\ T^{-3} + {\rm O}(T^{-4})$ while the Berry
parameter is $\delta_{\rm B} = \delta_{\infty} + b\ T^{-1} + {\rm
  O}(T^{-2})$ with $\delta_{\infty} = 0.3768(1)$. Here $a$ and $b$ are
volume and particle number specific constants. An analytical scaling
analysis reveals that $\delta_{\infty}$ depends on the container
geometry but neither on its volume nor the particle number.

\section{Dissociation}\label{sectionDiss}

Between the low temperature liquid phase and the
high temperature ideal gas limit Fig.\ \ref{CRofLargeT} shows clear
maxima both in the specific heat and the Berry parameter for Al$_2$,
Al$_7$, as well as for Al$_{13}$. The feature is generic for all
cluster sizes and can be associated with the dissociation
transition. The dissociation anomaly in the specific heat stems from
the increase in entropy across the dissociation transition while the
Berry parameter is enhanced through the short-time elongation and
return of an atom from and to the cluster. The latter involves an
energy fluctuation and is consequently suppressed in a microcanonical 
or isokinetic ensemble. 

\begin{figure}
     \resizebox{0.485\textwidth}{!}{%
       \includegraphics*[clip=true]{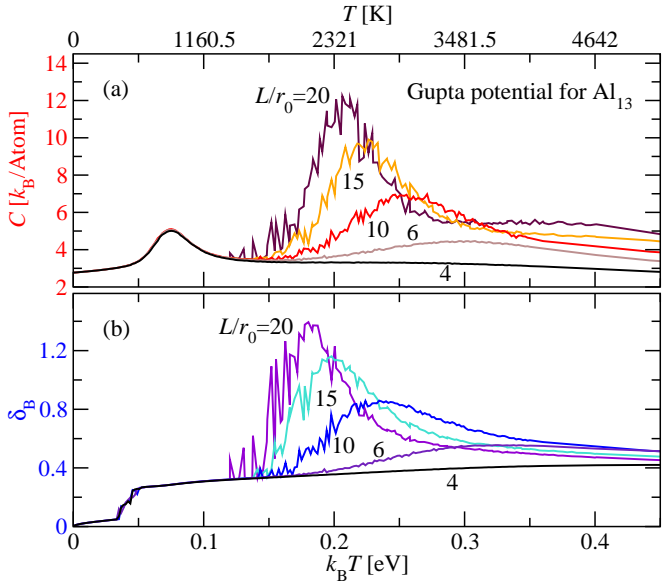}}
     \caption{\label{CRofSize}\sl
     Sample volume dependence of the dissociation transition for four
     different sampling volumes of $L^3/r_0^3 = 4^3$, $6^3$, $10^3$,
     $15^3$, $20^3$. Both the anomaly in the specific heat (upper 
     panel) and the Berry parameter (lower panel) increase with
     increasing volume as a consequence of the increased phase
     space. SR: $2\times10^7$ ($L^3/r_0^3 = 4^3$, $6^3$, 
     $10^3$) and $4\times10^7$ ($L^3/r_0^3 = 15^3$, $20^3$) steps per
     temperature and atom.    
   }
   \end{figure}

\begin{figure}
  \resizebox{0.485\textwidth}{!}{%
    \includegraphics*[clip=true]{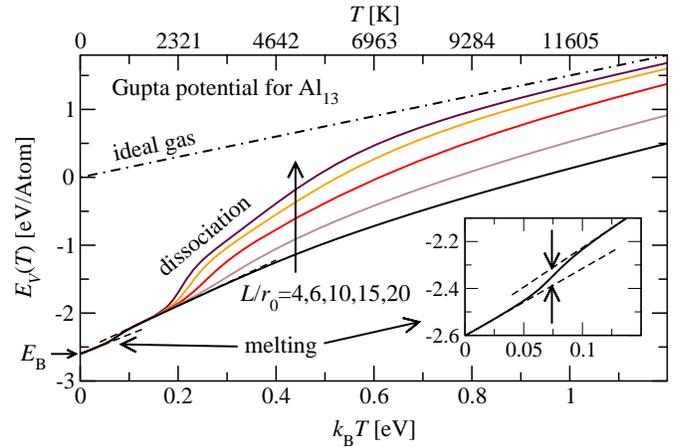}}
  \caption{\label{caloric13}\sl
    Caloric curves as obtained from integrating the specific heat data
    in Fig.\ \protect\ref{CRofSize} for sampling volumes of
    $L^3/r_0^3 = 4^3$, $6^3$, $10^3$, $15^3$, $20^3$. The inset shows the
    determination of the latent heat $\Delta E_{\rm melt} \approx
    0.08$ eV of the melting transition. The dash-dotted line is the
    ideal gas case reached only for temperatures much larger than the
    binding energy, i.e., $k_{\rm B}T \gg 2.60088$ eV. 
  }
\end{figure}

Figure \ref{CRofSize} shows the container size dependence of the
dissociation transition of an Al$_{13}$ cluster for volumes of $L^3 =
(4r_0)^3$, $(6r_0)^3$, $(10r_0)^3$, $(15r_0)^3$, and $(20r_0)^3$. For
$L^3 = (4r_0)^3$ the density $N/L^3 = 0.0144$ mole/cm$^3$ is only a
factor $7$ smaller than that of bulk Al with $0.1$ mole/cm$^3$. This
leads essentially to a suppression of the evaporation transition,
which is replaced by a smooth crossover (see also Fig.\
\ref{caloric13}). Note that even macroscopic particles do not exhibit
a sharp evaporation transition in a finite, constant volume imposing a
finite, temperature dependent gas-phase partial pressure. As a
consequence there is a liquid-vapour coexistence region which is given
by the width of the specific heat anomalies shown in Fig.\
\ref{CRofSize}. The snapshot shown in Fig.\ \ref{Tvapor}(b) in Sec.\
\ref{fragmenation} is taken in the liquid-vapour coexistence region of
Al$_{13}$ for $L^3 = (20r_0)^3$, where liquid fragments of the cluster
coexist with evaporated single atoms.

Both the anomaly in the specific heat (upper panel 
in Fig.\ \ref{CRofSize}) and the Berry parameter (lower panel in Fig.\
\ref{CRofSize}) increase with increasing volume and become
narrower. In an infinite volume the partial pressure of the gas phase
is zero and both quantities are expected to exhibit a sharp peak at
the transition. Note that the fluctuations of the graphs in Fig.\
\ref{CRofSize} increase with increasing volume as a consequence of the
enlarged phase space and thus limit the simulations to small
volumes. The melting transition is independent of the container
volume. This is expected since the pressure in the system is
essentially zero when all atoms are condensed \cite{SHD+01}.

For sufficiently small densities or, equivalently, large volumes the
latent heat of the melting transition is much smaller than the energy
released in the evaporation transition \cite{SHD+01}. For Al$_{13}$
this becomes apparent from the caloric curves of the total enclosed
system shown in Fig.~\ref{caloric13}, which are obtained by simply
integrating the specific heat (Fig.~\ref{CRofSize}) for a fixed
volume, i.e., 
$
E_V(T) = \int_0^T C(T')dT' - E_{\rm B}. 
$
The binding energy per atom for $N=13$ is $E_{\rm B} = 2.60088$
eV. Since the transitions are smeared out the determination of the
latent heat is  somewhat ambiguous. Extrapolation of the linear
segments of the caloric curves above and below the melting transition
(dashed lines in Fig.~\ref{caloric13}) and reading off the energy
difference at the temperature of the specific heat maximum yields
$\Delta E_{\rm melt} \approx 0.08$ eV.

The evaporation transition is very broad for the volumes (densities)
under investigation and the ideal gas limit ($E_{\rm ideal}(T) = 2/3
k_{\rm B} T$, dash-dotted line in Fig.\ \ref{caloric13}) is only
attained for $k_{\rm B}T > 1$ eV even for $L/r_0 = 20$. For $L/r_0 =
4$ the ideal gas limit is reached only for temperatures much larger
than the binding energy per atom, i.e., $k_{\rm B}T \gg E_{\rm B}$. 
Consequently the total latent heat of the evaporation can only be
given approximately as $\Delta E_{\rm evap} \approx 2.26$ eV, which
corresponds satisfactorily to the potential energy expectation value
$\langle V \rangle(T_{\rm evap}) = 2.35$ eV near the onset of the
evaporation transition at $k_{\rm B}T_{\rm evap} = 0.15$ eV.

Results for Al$_{55}$ (not shown) are similar with slightly narrower
evaporation anomalies compared with those of Al$_{13}$ with comparable 
densities $N/L^3$. Closed shell effects do not play an observable role
in the evaporation transition.  

From the above discussion follows that the dissociation anomalies in
the specific heat usually dominate in size over those of the melting
transition especially for clusters that do not have closed-shell
structures. An example is Al$_7$ as shown in the middle panel of Fig.\
\ref{CRofLargeT}. The jump in $\delta_{\rm B}$ is yet well pronounced
confirming its sensitivity to the melting transition \cite{ZKBB02}.

\section{Fragmentation}\label{fragmenation}

On closer examination the curves of the specific heat of Al$_{13}$ for
sufficiently large volumes show another broad local maximum at higher
temperatures. For $L/r_0 = 20$ as shown in Fig.\ \ref{CRofSize}(a)
this maximum is found near $k_{\rm B}T \sim 0.36$ eV. This effect is
readily explained by the fragmentation of the cluster into dimers and
trimers at temperatures above the onset of the evaporation transition
near $k_{\rm B}T_{\rm evap} = 0.15$ eV. The dimers and trimers in turn
dissociate at temperatures near the second local maximum. For smaller
volumes (larger densities) the effect is not visible because the
signatures become very broad. 

\begin{figure}
  \resizebox{0.485\textwidth}{!}{%
    \includegraphics*[clip=true]{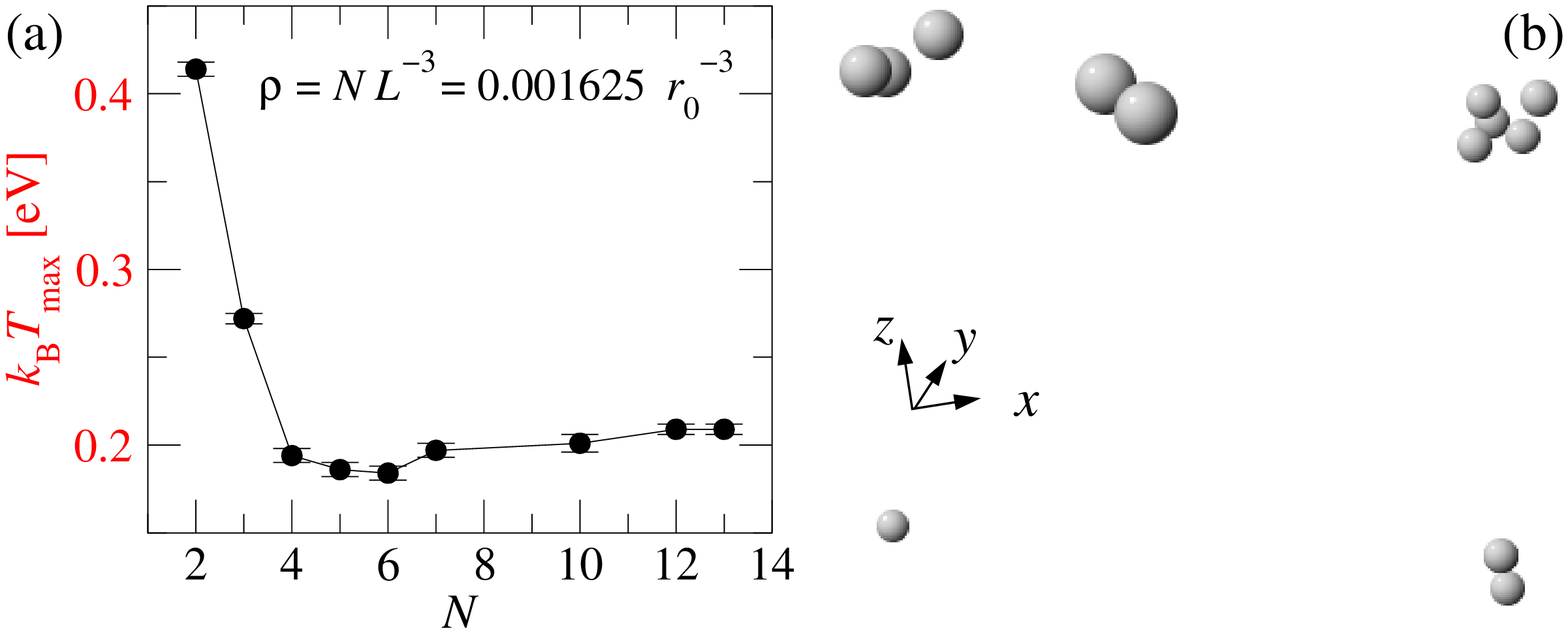}}
  \caption{\label{Tvapor}\sl
    (a) Temperature of the maximum of the specific heat at the evaporation
    anomaly as a function of the cluster size at a constant density
    of $\rho = 1.625\times 10^{-3} r_0^{-3}$. The line is a guide to
    the eye. The minimum at $N=6$ implies that those clusters
    dissociate first while trimers and dimers are stable to much
    higher temperatures. Larger clusters dissociate predominantly into
    dimers and trimers at intermediate temperatures as shown by the
    snapshot in panel (b) for Al$_{13}$ at $k_{\rm B}T = 0.25$ eV. SR
    for the determination of $T_{\rm max}$: $10^8$ steps per
    temperature and atom.  
  }
\end{figure}

To quantify the effect, Fig.\ \ref{Tvapor}(a) shows the maxima of the
evaporation anomaly of the specific heat as a function of the cluster
size at constant densities of $\rho = N/L^3 = 1.625\times 10^{-3} r_0^{-3}$,
which correspond for $N = 13$ to $L/r_0 = 20$ as shown in Fig.\
\ref{CRofSize}(a). The temperature of the specific heat maximum
increases with increasing size for $N \ge 6$ but for $N \le 6$ shows a 
non-monotonous {\em decrease} with increasing cluster size. In other
words, especially the dimers and trimers are more stable with respect
to larger clusters. The superposition of their specific heat maxima at
$k_{\rm B}T_{\rm max} = 0.415(5)$ eV and $k_{\rm B}T_{\rm max} =
0.275(5)$ eV for the dimer and trimer, respectively, leads to the
second maximum of the curve for $L/r_0 = 20$ in Fig.\
\ref{CRofSize}(a). Fig.\ \ref{Tvapor}(b) shows a snapshot of Al$_{13}$
at $k_{\rm B}T = 0.25$ eV with $L/r_0 = 20$, where a dissociated
single atom, two dimers and a trimers are visible together with a
(liquid) pentamere. Qualitatively comparable results are found for
Al$_{55}$ (not shown) with $k_{\rm B}T_{\rm max} = 0.23(1)$ eV at
similar densities.

Note that the stability of the dimers and trimers towards dissociation
is an entropic effect since the binding energy per atom decreases
monotonously with decreasing size for $N \le 13$
\cite{Binding}. Consequently the observed presence of dimers and
trimers is not in contradiction with the results from density
functional theory calculations \cite{RJ99}, which find the single atom
emission to be energetically the dominant dissociation channel. The
dimers and trimers are at least in part formed by fusion of single
evaporated atoms. On the other hand, this fragmentation behaviour is
not directly transferable to experiments on Al clusters because the
Gupta potentials do not yield the correct planar structure for $N \le
5$ \cite{AE99,RJ99}.

\section{Conclusions}

The results from the Monte-Carlo simulation of Al$_N$ clusters
modelled with many-body GPs reveals the appearance of a distinct
feature of the melting transition in the specific heat and the Berry
parameter for $N\ge6$. The energy fluctuations in the canonical
ensemble lead to an onset of isomer fluctuations at lower temperatures
than in the microcanonical ensemble. Al$_N$ clusters with $N\ge 4$ are
liquid at the bulk melting temperature. The present analysis suggests
that a generalised sufficient criterion for clusters not to be
considered solid any more is a Berry parameter of $\delta_{\rm B} \ge
0.1$. Larger clusters with closed-shell configurations exhibit sharper 
signatures of the melting transition than others.  

For higher temperatures clusters of all sizes undergo a dissociation
transition which is accompanied by container volume dependent
anomalies both in the specific heat and Berry parameter. Dimers and
trimers are more stable towards dissociation than larger clusters. The
details of the features depend on the potential but the qualitative
results are generic. For example, MC simulations with GPs and
parameters for Au yield qualitatively very similar results.


The author thanks M.\ Blom, P.\ Schmitt\-eckert, G.\ Schneider, D.\
Schooss, M.\ Vojta, and P.\ W\"olfle for instructive discussions. The
work was supported by the Center for Functional Nano\-struc\-tures of
the Deutsche Forschungsgemeinschaft within project D1.5.


\end{document}